\documentclass[12pt]{article}
\begin{document}

\newcommand{\ba}{\begin{array}}
\newcommand{\ea}{\end{array}}
\newcommand{\be}{\begin{equation}}
\newcommand{\ee}{\end{equation}}
\newcommand{\bea}{\begin{eqnarray}}
\newcommand{\eea}{\end{eqnarray}}
\newcommand{\beas}{\begin{eqnarray*}}
\newcommand{\eeas}{\end{eqnarray*}}

\parskip 10 pt

\begin{titlepage}

\begin{flushright}
CU-TP-1139 \\
hep-th/0511226 \\
\end{flushright}

\vskip 1.2 true cm

\begin{center}
{\Large \bf A note on the Coulomb branch of susy Yang-Mills}
\end{center}

\vskip 0.6 cm

\begin{center}

Daniel Kabat${}^1$ and Gilad Lifschytz${}^2$

\vspace{3mm}

${}^1${\small \sl Department of Physics} \\
{\small \sl Columbia University, New York, NY 10027 USA} \\
\smallskip
{\small \tt kabat@phys.columbia.edu}

\vspace{3mm}

${}^2${\small \sl Department of Mathematics and Physics} \\
{\small \sl University of Haifa at Oranim, Tivon 36006 ISRAEL} \\
\smallskip
{\small \tt giladl@research.haifa.ac.il}

\end{center}

\vskip 0.8 cm

\begin{abstract}
\noindent
We compute the force between oppositely charged $W$ bosons in the
large $N$ limit of Yang-Mills with 16 supercharges broken to $SU(N)
\times U(1)$ by a finite Higgs vev.  We clarify some issues regarding
Wilson line computations and show that there is a regime in which the
force between $W$ bosons is independent of separation distance.
\end{abstract}

\end{titlepage}

\section{Basic framework}

We first recall the computation, in the context of AdS/CFT duality
\cite{ads}, of the energy of a pair of $W$ bosons from supergravity
\cite{ry,maldaw}.  The background metric describing the near-horizon
geometry of $N$ coincident D$p$-branes is \cite{imsy}
\begin{eqnarray}
\nonumber
ds^2&=&\alpha'\left[\frac{U^{(7-p)/2}}{d_{p}^{1/2}e}(-dt^{2} 
+dx_{||}^2)+\frac{d_{p}^{1/2}e}
{U^{(7-p)/2}}(dU^2+U^2d\Omega^{2}_{8-p})\right]\\
\label{background}
e^{\phi}&=&(2\pi)^{2-p}g_{\rm YM}^{2}\left(\frac{d_{p}e^2}{U^{7-p}}\right)^{\frac{3-p}{4}}\\
\nonumber
d_{p}&=&2^{7-2p}\pi^{\frac{9-3p}{2}}\Gamma\Bigl(\frac{7-p}{2}\Bigr)\,.
\end{eqnarray}
Here $e^2 = g^2_{\rm YM} N$ is the 't Hooft coupling of the dual gauge
theory.  We place a probe D$p$-brane at some position $U_f$ and look
for a static solution describing a fundamental string that starts and
ends on the probe brane.  The string equations of motion follow from
the action
\begin{eqnarray}
\label{Stotal}
&&S = S_{\rm bulk} + S_{\rm bdy} \\
\nonumber
&&S_{\rm bulk} = -\frac{1}{2\pi\alpha'}\int d\tau d\sigma \sqrt{-\det G_{\mu\nu}
\partial_{m}X^{\mu}\partial_{n}X^{\nu}} \\
\label{Sbdy}
&&S_{\rm bdy} = \oint ds\, A_{\mu}\partial_{s}X^{\mu}
\end{eqnarray}
We have included a boundary action to allow for a $U(1)$ field
strength on the probe brane.

The simplest ansatz for a static string is
\begin{equation}
X^0 =\tau, \ \ \ X^1=\frac{L}{\pi}\sigma,\ \ \,U = U(\sigma)
\end{equation}
where the worldsheet coordinates range over $-\infty < \tau < \infty$,
$0\leq \sigma \leq \pi$.  Evaluated on the ansatz the bulk action
becomes
\begin{equation}
S_{\rm bulk}=-\frac{1}{2\pi}\int d\tau d\sigma \left(U'{}^{2}+\frac{L^2 U^{7-p}}
{\pi^2 d_{p}e^2}\right)^{1/2}\,.
\end{equation}
The equations of motion are solved by a U-shaped curve symmetric
around $\sigma=\pi/2$.  For $\pi/2<\sigma<\pi$ the solution is
\begin{equation}
\label{solution}
\sigma = {\pi \over 2} + \frac{\pi d_{p}^{1/2}e}{LU_{0}^{(5-p)/2}}\int_{1}^{U/U_{0}}\frac
{dy}{y^{(7-p)/2}(y^{7-p}-1)^{1/2}}\,.
\end{equation}
Here $U_{0} \equiv U(\pi/2)$.  Setting $U(\pi) = U_f$ fixes the
relationship between $L$ and $U_{0}$,
\begin{equation}
L=\frac{2d_{p}^{1/2}e}{U_{0}^{(5-p)/2}}\int_{1}^{U_f /U_0}
\frac{dy}{y^{(7-p)/2}(y^{7-p}-1)^{1/2}}\,.
\label{L}
\end{equation}
The string endpoints do not obey Neumann boundary conditions in the
coordinate $X^{1}$ \cite{kl}.  Denoting $\partial_{n},\partial_{s}$
the unit normal and tangential derivatives on the worldsheet, one
finds that
\begin{equation}
\partial_{n}X^1 = 2\pi\alpha' F_0{}^1 \partial_{s}X^{0}
\label{bc}
\end{equation}
where $F_{0}{}^{1}={1 \over 2\pi\alpha'}(U_0 / U_f)^{(7-p)/2}$.  This
corresponds to the inclusion of the boundary term in the action
(\ref{Sbdy}), and fixes the value of the electric field which must be
introduced on the probe brane.

The gravitational system we have described is dual to a Yang-Mills
theory with 16 supercharges in $p+1$ dimensions, with a finite Higgs
vev that breaks to $SU(N) \times U(1)$ and generates a $W$ mass $m_W =
U_f / 2\pi$.  The behavior of this system as $U_f \rightarrow \infty$
was analyzed in \cite{bisy}; for a treatment of the $U_f \rightarrow
\infty$ limit of breaking to $SU(N)\times SU(N)$ see \cite{mw}.  The
U-shaped fundamental string is dual to a pair of $W$ bosons,
oppositely charged under the unbroken gauge group, and $F_{0}{}^{1}$
is related to the $U(1)$ electric field one has to turn on for the
configuration to be static (it counterbalances the attraction from
$SU(N)$ interactions).  In the dual Yang-Mills the $U(1)$ electric
field is
\begin{equation}
E = F_{01} = \frac{U_{0}^{(7-p)/2}}{2\pi d_{p}^{1/2}e}.
\end{equation}
Note that the electric field is non-zero even if $U_f \rightarrow
\infty$.  Had we set $U_f$ to infinity from the beginning it would have
been hard to see the electric field, since in this limit it naively
looks as though $X^1$ obeys Neumann boundary conditions.

The attractive force $F$ on each $W$ boson due to the $SU(N)$
interactions exactly balances the force due to the electric field, so
we can identify\footnote{The U(1) charges at the string endpoints are
  $\pm 1$ in the normalizations used in (\ref{Sbdy}) and (\ref{bc}).}
\[
F = - E = - \frac{U_{0}^{(7-p)/2}}{2\pi d_{p}^{1/2}e}\,.
\]
Evaluating this as a function of the separation distance $L$ requires
inverting (\ref{L}) to find $U_0 = U_{0}(L)$.

Evaluating the string action (\ref{Stotal}) on the solution (\ref{solution})
one finds the bulk and boundary contributions
\begin{eqnarray}
\nonumber
&&S_{\rm bulk} = \int d\tau \, - \frac{U_0}{\pi}\int_{1}^{U_f/U_0}\frac{dy y^{(7-p)/2}}
{(y^{7-p}-1)^{1/2}} \\
&&S_{\rm bdy} = \int d\tau  L E\,.
\label{sb}
\end{eqnarray}
In the dual gauge theory $S_{\rm bulk}$ gets identified with the
effective action for the $SU(N)$ sector of the dynamics, while $S_{\rm
bdy}$ is the effective action for the $U(1)$ sector.\footnote{The
fact that the effective action decomposes in this way is a
consequence of large $N$.}  For a static configuration the
Lagrangian is minus the Hamiltonian, so we can identify the potential
energy in the $SU(N)$ sector (which includes the rest mass of the $W$
bosons) as
\begin{equation}
V = \frac{U_0}{\pi}\int_{1}^{U_f/U_0}\frac{dy y^{(7-p)/2}}
{(y^{7-p}-1)^{1/2}}\,.
\end{equation}
The integrals we've encountered evaluate to hypergeometric functions.
Defining $z=(U_f/U_0)^{7-p}-1$ the potential and separation distance
are
\begin{eqnarray}
\label{potential}
V&=&\frac{2U_{0}}{(7-p)\pi}z^{1/2}\,{}_2F_{1}\left(\frac{5-p}{2(7-p)},\frac{1}{2},
\frac{3}{2},-z\right) \\
\label{fe}
L&=&\frac{4d_{p}^{1/2}e}{(7-p)U_{0}^{(5-p)/2}}z^{1/2}
\,{}_2F_{1}\left(\frac{5-p}{2(7-p)}+1,\frac{1}{2},\frac{3}{2},-z\right)\,.
\end{eqnarray}
With a bit of work one can check that $F = - {\partial V
\over \partial L}$; the identities
\begin{eqnarray}
&&z\frac{d}{dz}\,{}_2F_{1}(a,b,c,z)-a\,{}_2F_{1}(a+1,b,c,z)+a\,{}_2F_{1}(a,b,c,z)=0 \\[3pt]
\nonumber
&&(2a-c+z(b-a))\,{}_2F_{1}(a,b,c,z)+(c-a)\,{}_2F_{1}(a-1,b,c,z) \\
&&\quad+a(z-1)\,{}_2F_{1}(a+1,b,c,z)=0
\end{eqnarray}
are useful.

At this stage we can already see that the force should exhibit some
interesting behavior.  Due to the Born-Infeld action on the probe
brane \cite{ft,acny,lei} there is a limiting value for the electric
field.  So the force must be bounded, even as $L \rightarrow 0$.  We
will see that this is indeed the case.

\section{Scales and range of validity}

There are several restrictions on the validity of our results.  The
first is that the supergravity background must be trustworthy: the
entire string worldsheet must be in a region in which the curvature is
small.\footnote{If the dilaton becomes large we can always go to an
S-dual or M-theory description \cite{imsy}.}  The radius of
curvature of the background (\ref{background}) is $R_{\rm curvature} \sim
\ell_s e^{1/2} / U^{(3-p)/4}$.  Requiring that this exceed the string
length at the position of the probe brane gives the condition
\begin{equation}
\label{SugraValid}
U_{f}^{3-p} \ll e^2.
\end{equation}
For $p<3$ this means we can not send the probe brane to infinity.  For
$p>3$ we should also impose the stronger condition
\begin{equation}
U_0^{3-p} \ll e^2
\end{equation}
which gives an upper bound on $L$.

Even if the supergravity background is reliable we still need to make
sure that we can treat the string configuration classically.  Denote
the proper distance between the two string endpoints (measured along
the probe brane) by $L_{\rm proper}$.  Although we do not know how to
quantize the string it seems reasonable to demand that $L_{\rm
proper}$ is larger than the string length, $L_{\rm proper} \gg
\ell_s$.  This gives the condition
\begin{equation}
\frac{L U_{f}^{(7-p)/4}}{\sqrt{e}} \gg 1.
\label{conls}
\end{equation}
At large 't Hooft coupling this is a much stronger condition than the
requirement that $L_{\rm proper}$ be larger than the $W$ Compton
wavelength $m_W^{-1} = 2 \pi / U_f$.  In the large $N$ limit it's also
much stronger than the requirement that $L_{\rm proper}$ exceed the
Planck lengths
\[
\ell_{10} = {\ell_s \over N^{1/4}} \biggl({e \over U_f^{(3-p)/2}}\biggr)^{(7-p)/8} \qquad\quad
\ell_{p+2} = {\ell_s \over N^{2/p}} \biggl({e \over U_f^{(3-p)/2}}\biggr)^{(6-p)/2p}
\]
in 10 and $p+2$ dimensions, respectively.

We will see that the force between $W$ bosons exhibits interesting
behavior when $\ell_s \ll L_{\rm proper} \ll R_{\rm curvature}$, or
equivalently when
\begin{equation}
\label{conint}
{\sqrt{e} \over U_f^{(7-p)/4}} \ll L \ll {e \over U_f^{(5-p)/2}}\,.
\end{equation}
Note that this can be satisfied only if supergravity is valid,
$U_f^{3-p} \ll e^2$.  In the AdS/CFT context one usually relates
localized objects in the bulk to delocalized excitations on the
boundary.  The UV/IR correspondence \cite{suswit,pp} implies that in
the regime (\ref{conint}) the excitations representing the two $W$'s
will overlap on the boundary.  One may then wonder about the validity
of the computation: since the excitations overlap, have we overlooked
something?  We believe the answer is no: in the large $N$ limit there
is no reason to think the computation is not valid.  For example, in
the semiclassical limit local bulk excitations can be represented by
smeared operators on the boundary.  However bulk supergravity
correlation functions are precisely reproduced by correlators in the
boundary theory, even when the smeared operators completely overlap
\cite{hkll1,hkll2}.

\section{Examples}

\subsection{D3-branes}

We first consider the case $p=3$.  Since we are working at large 't
Hooft coupling, the condition for validity of the supergravity
background (\ref{SugraValid}) places no restrictions on $U_f$ or $L$.
The force between the $W$ bosons is
\begin{equation}
F = -\frac{U_{0}^{2}}{2\sqrt{2}\pi e}
\end{equation}
while the separation distance is
\begin{eqnarray}
\label{D3sep}
L = \frac{2\sqrt{2}e}{U_0}\int_{1}^{U_f /U_0}
\frac{dy}{y^{2}(y^{4}-1)^{1/2}}\,.
\end{eqnarray}
This can be evaluated in terms of the first and second elliptic
integrals ${\cal{F}},{\cal{E}}$.
\begin{equation}
L = \frac{2e}{U_{0}}\Bigl(2{\cal{E}}\bigl(\arccos (U_0/U_f),1/\sqrt{2}\bigr) -
{\cal{F}}\bigl(\arccos (U_0/U_f),1/\sqrt{2}\bigr)\Bigr)
\end{equation}

Following \cite{ry,maldaw} let's first study the behavior for
large $U_f$ (large $W$ mass).  As $U_f \rightarrow \infty$ we have
\begin{equation}
L = \frac{4\pi^{3/2}e}{U_0\Gamma^{2}(1/4)}\,.
\end{equation}
Then as a function of the separation distance, the force between the $W$
bosons is
\begin{equation}
F=-\frac{4\pi^{2}\sqrt{2}e}{\Gamma^{4}(1/4)}\,{1 \over L^2}
\end{equation}
while the potential (\ref{potential}) behaves as
\begin{equation}
V = {U_f \over \pi} - \frac{4\pi^{2}\sqrt{2}e}{\Gamma^{4}(1/4)}\,\frac{1}{L}\,.
\end{equation}
The first term can be identified with the mass of the $W$ bosons, $2
m_W = U_f / \pi$, while the second term is the energy due to $SU(N)$
interactions.  The Coulomb-like $1/L$ behavior of the second term is
required by conformal invariance.

Now let's study the leading corrections to these results when $U_f$ is large but
finite.  Expanding (\ref{D3sep}) for large $U_f/U_0$ the separation
distance is
\begin{equation}
L = {2\sqrt{2}e \over U_0} \left({(2\pi)^{3/2} \over 2 \Gamma^2(1/4)} - \frac{1}{3}\Bigl(
\frac{U_0}{U_f}\Bigr)^3\right)
\end{equation}
and the force as a function of separation is
\begin{equation}
F = -\frac{4\pi^{2}\sqrt{2}e}{L^2 \Gamma^{4}(1/4)}
\left(1-{8\sqrt{2} (2\pi)^3 \over 3 \Gamma^4(1/4)}\Bigl(\frac{e}{L U_f}\Bigr)^3\right)\,.
\label{f1}
\end{equation}

Finally let's see what happens when $U_f/U_0$ is close to one.  Then
we can approximate
\begin{equation}
U_{0} = U_f \left(1-\frac{L^2 U_f^2}{8e^2}\right)
\end{equation}
to find that the force between the $W$ bosons is
\begin{equation}
F= - \frac{U_{f}^{2}}{2 \pi \sqrt{2} e}\left(1-\Bigl(\frac{L U_{f}}{2e}\Bigr)^2\right)\,.
\label{f2}
\end{equation}
This result applies in the regime (\ref{conint}), namely when
$\sqrt{e} \ll \ L U_f \ll e$.  Note that in this regime the force
between the $W$'s is roughly independent of separation.

\subsection{General $p$}

We first study the behavior for $U_f/U_0 \gg 1$.  We can approximate
the hypergeometric function in (\ref{fe}) to find ($B \equiv
B\left(\frac{1}{2},\frac{6-p}{7-p}\right)$ is the beta function)
\begin{equation}
F=-\frac{1}{2\pi d_{p}^{1/2} e}\left(\frac{2 B d_{p}^{1/2} e}
{(7-p)L}\right)^{\frac{7-p}{5-p}}
\left[1-\frac{B^{\frac{7-p}{5-p}}}{(5-p)(6-p)(7-p)^{2/(5-p)}}
\Bigl(\frac{2 d_{p}^{1/2} e}
{U_{f}^{(5-p)/2} L}\Bigr)^{\frac{12-2p}{5-p}}\right]
\end{equation}
The restrictions on $L$ are practically the same as in the $U_f
\rightarrow \infty$ limit studied in \cite{bisy}. For $p<3$ one can
always go to larger $L$ by using either an S-dual description or an
M-theory lift.

In the regime (\ref{conint}) $U_f / U_0$ is close to one.  Then we can
approximate
\begin{equation}
U_{0}=U_{f}\left(1-(7-p)\frac{L^2 U_{f}^{5-p}}{16 d_{p} e^2}\right)
\end{equation}
and the force is given by
\begin{equation}
F=-\frac{U_{f}^{(7-p)/2}}{2\pi d_{p}^{1/2}e}
\left(1-(7-p)^2\frac{L^2 U_{f}^{5-p}}{32 d_{p} e^2}\right)\,.
\end{equation}
Again we see a roughly constant force between the $W$'s when they are
separated by less than the curvature radius.

\subsection{D5-branes}

The case $p=5$ is a special as the integrals can be expressed in terms
of elementary functions.
\begin{eqnarray}
\nonumber
L & = & 2 d_5^{1/2} e \, \cos^{-1}(U_0/U_f) \\
\label{d5}
F & = & - \frac{U_{f}}{2\pi d_{5}^{1/2}e}\cos \frac {L}{2d_{5}^{1/2}e} \\
\nonumber
V & = & \frac{U_f}{\pi}\sin \frac {L}{2d_{5}^{1/2}e}
\end{eqnarray}
This system was studied in the limit $U_f \rightarrow \infty$ in
\cite{bisy}.  In this limit the force diverges unless $L$ is fixed.
But here we see that keeping $U_f$ finite leads to a reasonable
answer.
 
In the range (\ref{conint}) where $(e/U_f)^{1/2} \ll L \ll e$ equation
(\ref{d5}) shows a constant force between the $W$ bosons.  However we
should check whether the string enters a region where the curvature is
large.  The condition for small curvature at the tip is $U_0 e > 1$.
This places an upper bound on $L$,
\[
L < 2 d_{5}^{1/2} e \cos^{-1}\bigl({1 \over e U_f}\bigr) \approx \pi d_5^{1/2} e\,.
\]
Not surprisingly this is compatible with (\ref{conint}).

\section{Conclusions}

By introducing an electric field on the probe brane we have computed
the force between $W$ bosons at finite Higgs vev.  The new feature of
our results is that for separation distances $\ell_s \ll L_{\rm
proper} \ll R_{\rm curvature}$ the force between $W$ bosons is
independent of separation.  In a way this is no surprise: when $L_{\rm
proper} \ll R_{\rm curvature}$ the background can be approximated by flat
space, and classical strings give rise to a linear potential.

Our result is more surprising from the point of view of the dual gauge
theory.  Consider the case $p = 3$.  At large $U_f$ the gauge group is
broken at a high scale.  The $W$ bosons are very massive and
non-dynamical, so we have a conformally-invariant $SU(N)
\times U(1)$ gauge theory with an interquark potential $\sim 1/L$ as
required by conformal invariance.  As the Higgs vev is reduced the theory
begins to notice the broken scale symmetry associated
with the finite mass of the $W$ bosons.  At a critical Higgs vev
the form of the potential changes.  Equivalently, for fixed Higgs vev
the form of the potential changes at a critical separation distance.
Our results indicate that at large $N$ and large 't Hooft coupling
this occurs when the separation distance $L \sim e / U_f$.  This is
enhanced by a factor of the 't Hooft coupling compared to the distance
one might have naively expected, namely the Compton wavelength of the $W$
bosons $1/m_W = 2\pi / U_f$.  For discussion of a similar discrepancy
see \cite{Freedman:1999gk,Bianchi:2001de}.

\vskip 1.0 cm
\goodbreak
\centerline{\bf Acknowledgments}
We are grateful to Kostas Skenderis for comments on the manuscript.
This research was supported in part by the US-Israel Binational
Science Foundation under grant \#2000359.  DK is also supported by the
US Department of Energy, grant DE-FG02-92ER40699.

\end{document}